\documentclass[journal=nalefd,manuscript=letter]{achemso}

\usepackage[version=3]{mhchem} 
\usepackage[table]{xcolor}
\usepackage{ulem}

\author{Simone Finizio}
\affiliation{Paul Scherrer Institut, 5232 Villigen PSI, Switzerland}
\email{simone.finizio@psi.ch}
\author{Benjamin Watts}
\affiliation{Paul Scherrer Institut, 5232 Villigen PSI, Switzerland}
\author{Benedikt R\"osner}
\affiliation{Paul Scherrer Institut, 5232 Villigen PSI, Switzerland}
\author{Tim A. Butcher}%
\affiliation{Max Born Institute for Nonlinear Optics and Short Pulse Spectroscopy, 12489 Berlin, Germany}
\author{Sebastian Wintz}%
\affiliation{Helmholtz Zentrum Berlin f\"ur Materialien und Energie, 12489 Berlin, Germany}
\author{Markus Weigand}%
\affiliation{Helmholtz Zentrum Berlin f\"ur Materialien und Energie, 12489 Berlin, Germany}
\author{J\"org Raabe}
\affiliation{Paul Scherrer Institut, 5232 Villigen PSI, Switzerland}

\title{Supersampled scanning transmission X-ray microscopy for high-resolution vibration-independent time-resolved imaging}


\begin{document}




\begin{abstract}
Scanning transmission X-ray microscopy (STXM) is a nanoscale imaging technique that can utilize several powerful contrast mechanisms for the quantitative mapping of chemical and physical materials properties. Spatial resolutions down to 7~nm at the soft X-ray energy range have been demonstrated. A limiting factor for high-resolution STXM imaging is given by the positioning precision of the sample with respect to the focusing optic, with the current state-of-the-art leading to significant overheads, especially at low pixel dwell times, and being vulnerable to unavoidable external vibrations sources. In this work, we present a method, called \textit{supersampled scanning microscopy} that allows for a significant reduction of overhead times while simultaneously removing the effects of vibrational noise by sampling the position of the sample at a rate significantly higher than the vibration spectrum and reconstructing the sample transmission image from the recorded list of positions and detector counts. We demonstrate the performance of the technique with a set of proof-of-concept high-resolution imaging experiments.
\end{abstract}


\section{Introduction}

In recent decades, third-generation synchrotron lightsources have provided bright photon beams with photon energies ranging from infrared to hard X-ray wavelengths. The availability of highly brilliant X-ray radiation has allowed for countless scientific discoveries in a wide range of fields. Recently, a new fourth-generation of synchrotron lightsources has been developed, and several of these new ring designs have been built, mostly by upgrading existing rings. This next generation is referred to as diffraction limited storage rings (DLSRs), and offers significant increases in the source brilliance and the coherent photon flux delivered to the beamlines of typically one to two orders of magnitude compared to third-generation sources depending on the lattice type, source properties, and photon energies. The impact is expected to revolutionize scientific investigations with synchrotron light in many fields \cite{Streun2018, Raimondi2023}.

One portfolio of techniques that will profoundly benefit from the upgrade of synchrotron lightsources to DLSRs is X-ray microscopy, specifically photon-hungry techniques that require a high degree of coherent illumination such as nanofocus imaging or coherent diffractive imaging (CDI). A great example of the significant improvements offered by DLSRs is given by soft X-ray ptychography imaging, a CDI technique where the routine acquisition of high-resolution (sub-10 nm at the soft X-ray range) images was demonstrated \cite{Butcher2024, Butcher2025, Butcher2025b, Raimondi2023, Johansson2021, Walsh2021}. Besides CDI, direct nanofocus imaging techniques, such as scanning transmission X-ray microscopy (STXM), will strongly benefit from the increased beam quality. This technique utilizes a diffractive X-ray lens (Fresnel zoneplate - FZP) to focus a monochromatic X-ray beam onto a nanometric spot. In the case of STXM, the beam is focused on the surface of an X-ray semi-transparent sample and an image is obtained by scanning the sample and recording the transmitted X-ray intensity for each point of the scan. Under a sufficiently coherent illumination of the FZP, the achievable spatial resolution is limited by the diameter of the Airy disk generated by the FZP: the size of the central spot is 1.22 times as large as the outermost zone width of the FZP. Whereas it helps to decrease the outermost zone width to achieve higher microscopic resolution, two major limitations arise. One is the technological limitation of lithographic fabrication processes to produce the required FZP, and the other is the fact that high-resolution FZP usually suffer from decreasing diffraction efficiencies as volume effects kick in \cite{Schneider1997} and ever-smaller nanostructures also tend to be less perfect. The current soft X-ray STXM resolution record stands at 7~nm, achieved using a tailored zone-doubled Ir FZP \cite{Roesner2020}.

\begin{figure*}[hbt]
    \includegraphics[width=0.95\textwidth]{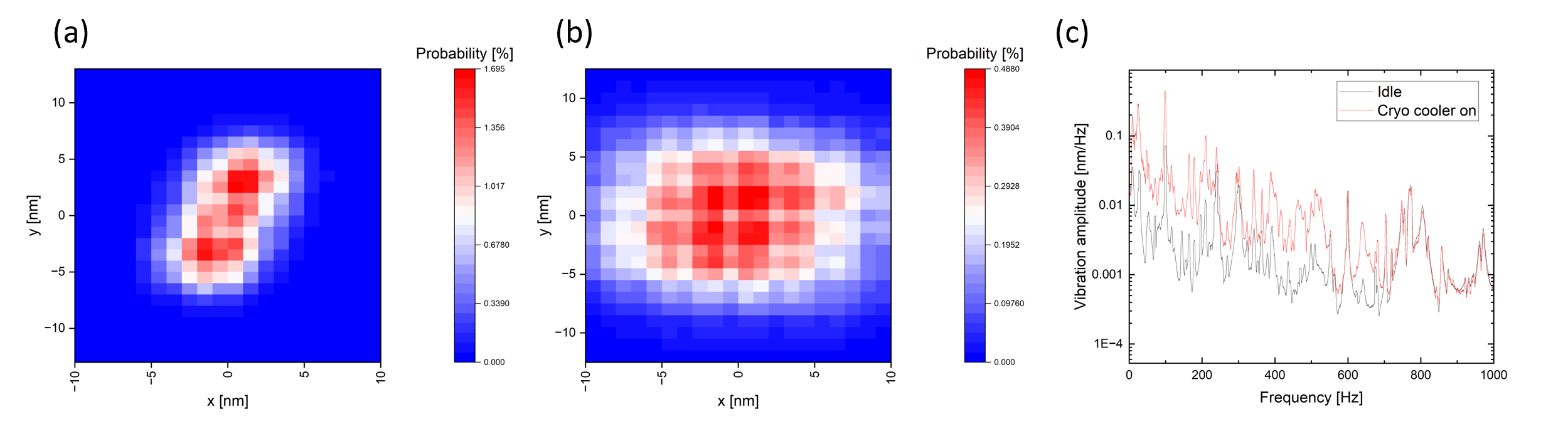}
    \caption{(a) Heatmap showing the sample positioning uncertainty in idling conditions at the PolLux endstation (control loop attempting to keep position); (b) Same conditions as (a), but with a closed cycle cryostat, normally employed for cooling down samples, activated; (c) Spectrum of the vibrations in both configurations.}
    \label{fig:vibrations}
\end{figure*}

While the principle of sub-10~nm resolution soft X-ray STXM imaging has been demonstrated already in 2020 \cite{Roesner2020}, its use in regular measurement campaigns has not been widespread, due to the combination of two preconditions: a relatively low efficiency of high-resolution FZP optics (as discussed above) and a very high positioning precision and stability. Further, the very short depth of focus of high resolution FZPs \cite{Roesner2020} means that the high positioning precision is required in all three spatial dimensions, although the level of precision required along the beam axis is still an order of magnitude less than the transverse precision requirements. When discussing positioning precision and stability, it is important to understand that the positioning stability does not need to be absolute, but relative to the time used to collect detector counts (i.e. the pixel dwell time). While DLSRs help to overcome the challenges for the focused beam size and the time required to collect sufficient photon counting statistics, this only solves half of the problem with high-resolution STXM imaging, as the reliable, stable, and high precision positioning of the sample with respect to the FZP is very difficult to achieve with conventional approaches.

In routine STXM imaging, sample positioning is performed by means of a piezoelectric stage with an interferometric feedback system. At the PolLux endstation of the Swiss Light Source \cite{Raabe2008} and at the Maxymus endstation of the Bessy II light source, the interferometric metrology is performed by a dual pass heterodyne interferometer \cite{Holler2015}. The difference between the recorded interferometer position and the desired position is then used in a proportional-integral-differential (PID) feedback loop to actuate the piezoelectric stage. This is performed using the Orocos Project libraries \cite{Orocos} on a realtime Linux kernel, which is operated with a 2~kHz clock. However, even with a fully optimized feedbacked control system, perfect positioning at the nanoscale cannot be achieved, due to the presence of environmental vibrations that cannot be completely canceled out. These vibrations lead to an uncertainty on the position of the probing beam on the sample surface and can therefore remarkably reduce the quality of the recorded STXM images. Fig.~\ref{fig:vibrations}(a) shows the uncertainty on the sample position at the PolLux endstation when the control loop is attempting to maintain the user-defined set position. A $1\sigma$ root mean square uncertainty of about 2.4~nm in the \textit{x} axis and of 3.8~nm in the \textit{y} axis can be observed. This is negligible for most applications of low-resolution imaging, where both spot size and scanning steps are much larger than the vibration amplitude, but it becomes relevant for high-resolution images when the X-ray spot size becomes comparable to the vibration amplitude. On top of this fundamental issue, the use of additional equipment in the endstation, such as e.g. a closed cycle cryostat used for sample cooling, can lead to a significant increase in the vibration amplitude (see Fig.~\ref{fig:vibrations}(b), with a significantly higher $1\sigma$ uncertainty of about 9.5~nm in the \textit{x} axis and 12.7~nm in the \textit{y} axis). By analyzing the spectrum of the vibrations (see Fig.~\ref{fig:vibrations}(c)) it is possible to observe that several vibration modes are present, with the strongest at approximately 100~Hz for this setup. For other STXM endstations, such as the Maxymus endstation at the Bessy II lightsource, the exact details of the positioning uncertainty and the vibrations' spectral content will vary, but the principle issue that environmental noise couples into the positioning will remain.

\begin{figure}[hbt]
    \includegraphics[width=0.4\textwidth]{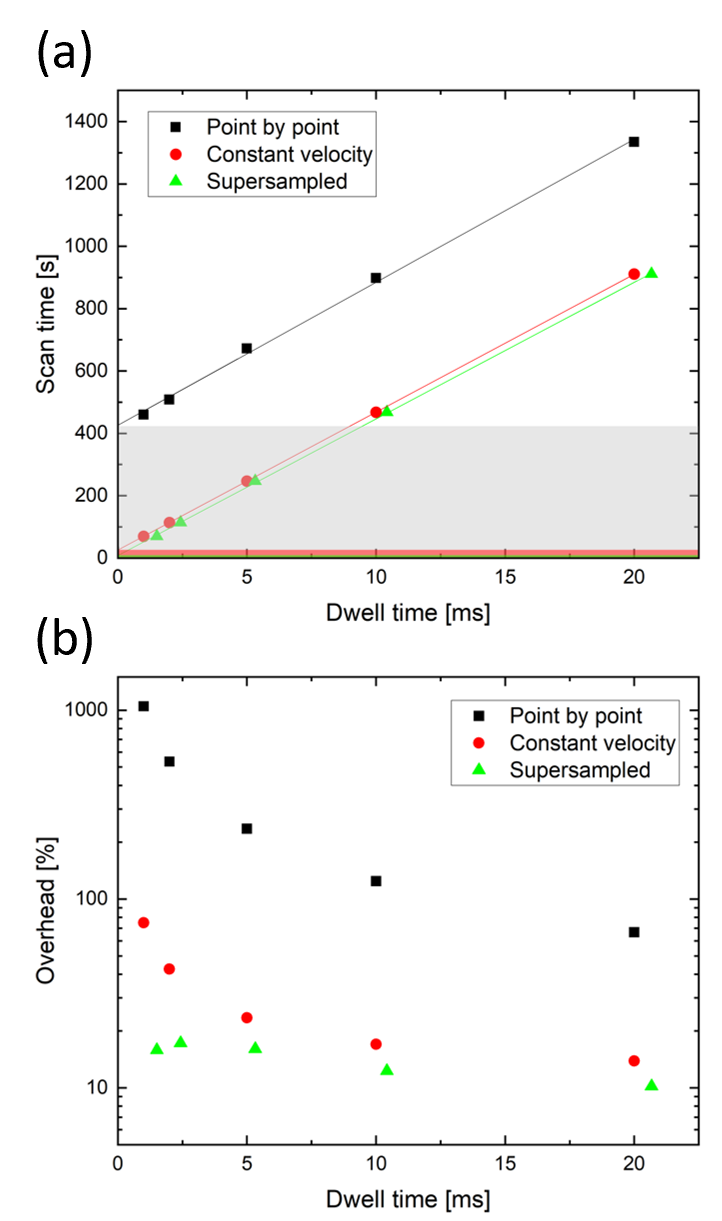}
    \caption{(a) Total scan time determined from STXM scans for a $200 \times 200$~px$^2$, $2 \times 2$~\textmu m scan in the PP, CV, and with the supersampled configurations as a function of the desired pixel dwell time. (b) Corresponding scan overhead fraction (calculated as the ratio between the overhead time and the nominal scan time).}
    \label{fig:overhead}
\end{figure}

The PolLux nd Maxymus STXM control systems utilize two scanning protocols: \textit{point-by-point} (PP) and \textit{constant velocity} (CV), also known as \textit{line-at-once}. The PP mode changes the setpoint of the positioning control loop for each pixel of the scan and waits for the position to settle before it records the transmitted photon intensity (by integrating the signal for a user-defined dwell time). The PP mode is precise within the uncompensated vibration noise, but at the price of large overheads required for the settling of the sample position at each pixel, which limits the scan speed. The second scanning protocol, the CV mode, trades positioning precision for speed. In this mode, each line of the region of interest (ROI) is scanned with the piezoelectric stage moving at a constant velocity, which is defined by the number of pixels in the line and the required dwell time (calculated as $v = d/(t_p \ N_p)$, $d$ being the length of the line, $t_p$ the pixel dwell time, and $N_p$ the number of pixels in the line). The transmitted photon intensity is continuously recorded and separated in $N_p$ bins, each with a $t_p$ duration in time. As the piezoelectric stage cannot instantaneously accelerate to or decelerate from a non-negligible velocity, a region before and after the ROI is dedicated to the acceleration and deceleration of the stage. This acceleration distance is user-defined, and depends on parameters such as the velocity to achieve and the dimension of the ROI. The acceleration distance contributes to the otherwise negligible overhead of CV scans, as part of the measurement time will be spent in the acceleration/deceleration regions. Fig.~\ref{fig:overhead} demonstrates the impact of the scanning protocol on the scanning overhead for an image, where the significant difference between the PP and the CV modes can be observed. The constant overhead weights more for shorter dwell times. With the increased brilliance of the source offered by DLSRs, shorter dwell times will become more common, and scanning overhead will therefore become more limiting under experimental conditions, which needs to be addressed.

In this work, we present a new STXM imaging protocol that allows for the combination of high-precision positioning for the acquisition of high-resolution images with negligible imaging overheads and the possibility to eliminate the effects of external vibrations, based on the idea of exploiting high-precision metrology to improve image quality in scanning probe microscopy \cite{Sun2021, Roesner2020, Holler2015}. This protocol is based on the principle that vibrations and positioning glitches can never be fully compensated for by conventional methods, but that they occur at timescales much longer than the interaction of the X-ray photons with the sample. By measuring the position of the sample at a rate faster than the mechanical vibrations, the true position at which the photon interacts with the sample can be determined with much higher precision and accounted for in the image reconstruction. In this way, unpredictable vibrations and positioning glitches simply become part of the scanning motions. In addition, this relaxes the need to achieve precise sample positions (as long as a sufficiently uniform coverage of the ROI can be provided), allowing for an increase in the scanning velocity to the physical limits of the stages.

\begin{figure}[hbt]
    \includegraphics[width=0.4\textwidth]{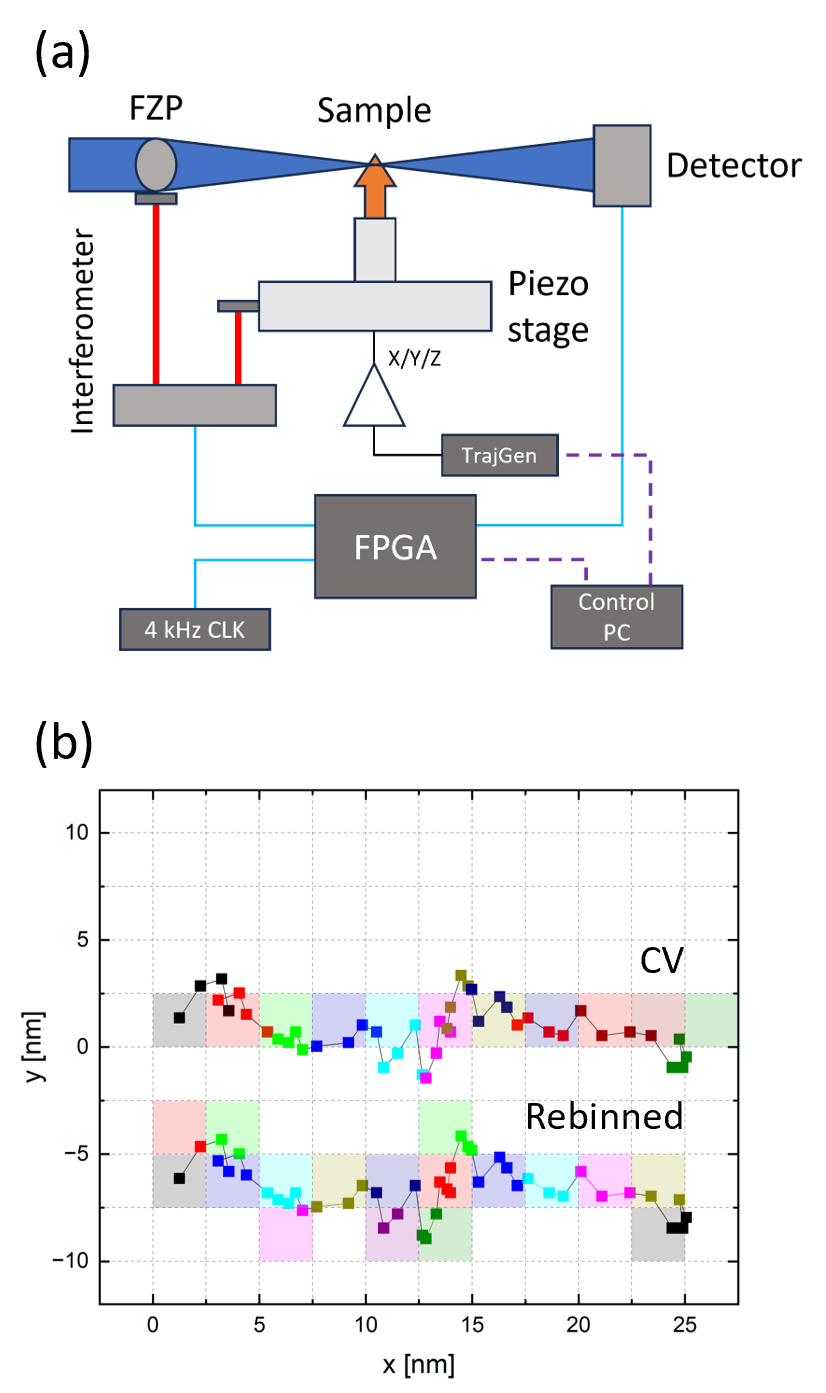}
    \caption{(a) Sketch of the setup employed for the supersampled scanning microscopy measurements. The position of the sample, together with the detected photon counts, is recorded at a fast (4~kHz) rate by a dedicated FPGA setup. The piezoelectric stage is scanned with minimal positioning precision requirements using a trajectory generator integrated within the Orocos positioning system of the endstation. (b) Sketch of the operation principle of the CV mode and of the principle of supersampled imaging for the same scan trajectory. In supersampled imaging, the data is assigned to a pixel based on the recorded position, allowing one to overcome possible artifacts caused by vibrations or glitches in the positioning system.}
    \label{fig:setup}
\end{figure}

The setup described in this work is sketched in Fig.~\ref{fig:setup}(a). The position of the sample with respect to the FZP is measured with a dual pass heterodyne interferometer, as described above. The PolLux interferometer (Zygo ZMI-4104) can be sampled up to about 770~kSa/s, while the Maxymus interferometer (Agilent 10719A, combined with an Agilent N1231B three-axis laser board) can be sampled up to about 20~MSa/s. In this work, the position is sampled with a 4~kHz clock using a dedicated field programmable gate array (FPGA) position sampling setup (PandABox, from Quantum Detectors \cite{Zhang2018}). For the PolLux setup, the encoder signal (a 32-bit number for both the \textit{x} and \textit{y} axes) is delivered to the PandABox through an asynchronous serial protocol proprietary of Zygo. For the Maxymus setup, the parallel output of the N1231B card (also delivering a 32-bit number for both the \textit{x} and \textit{y} axes) is pre-processed by a custom-designed FPGA board to convert the parallel output into the synchronous serial interface (SSI) signal accepted by the PandABox. The 4~kHz sampling clock is directly derived from the 125~MHz native clock of the PandABox FPGA, guaranteeing that the time stamps are synchronized with the detected photon counts. While higher sampling rates are possible, we selected a sampling rate of 4~kHz as a compromise between a reasonably low data stream rate (ca. 228~kB/s) while still sampling the majority of the mechanical vibration spectrum (see Fig.~\ref{fig:vibrations}(c)). For each clock cycle, the recorded photon counts from the detector are also saved. The recorded information is streamed by transmission control protocol (TCP) to a control computer that processes the data. The PandABox FPGA configuration used in this work is shown in the Supplementary Information. The scanning of the sample is performed by driving the piezoelectric stage amplifier using a trajectory generator implemented within the realtime kernel running the Orocos library, whereas the requirements on the positioning feedback have been significantly relaxed compared to the PP and CV modes.

For both implementations of the supersampled imaging setup at the PolLux and Maxymus endstations, no significant deviations from the usual stability of the operation of the STXM instrument or of the Pixelator control software were observed, indicating that the method can be easily implemented within existing setups.

\begin{table}[!h]
    \centering
    \begin{tabular}{c | c | c | c}
        $p_{\mathrm{M},x}$ [cts] & $p_{\mathrm{M},y}$ [cts] & Counts [cts] & Timestamp [s]\\
        \hline \hline
        ... & ... & ... & ...\\
        34858025 & 32360989 & 388513 & 107784.53560\\
        34858028 & 32360996 & 389071 & 107784.53585\\
        34858032 & 32361007 & 389616 & 107784.53610\\
        34858024 & 32361005 & 390193 & 107784.53635\\
        34858026 & 32361016 & 390757 & 107784.53660\\
        34858027 & 32361033 & 391338 & 107784.53685\\
        34858033 & 32361036 & 391913 & 107784.53710\\
        34858041 & 32361043 & 392464 & 107784.53735\\
        ... & ... & ... & ...\\
    \end{tabular}
    \caption{Example of a datastream from the PandABox. Note that the table shows only a subset omitting other diagnostic information for simplicity.}
    \label{tab:datastream}
\end{table}

An example of the datastream transmitted by the PandABox is depicted in Tab.~\ref{tab:datastream}. The datastream contains two columns, where the \textit{x} and \textit{y} interferometer positions are recorded at the rising edge of the 4~kHz clock. The position is recorded in resolution units of the interferometer, which corresponds to 150~pm/step for both PolLux and Maxymus. For experiments requiring 3D positioning, e.g. for tilted scans used to image in-plane magnetic systems \cite{Finizio2017}, or for laminography imaging \cite{Witte2020}, a third column recording the \textit{z}~interferometer position can be added. The next column contains the photon counts detected during each clock cycle. This is done by connecting the total transistor logic (TTL) output of the detector to one of the digital logic inputs of the PandABox that is linked to a counter in the FPGA. The counter is enabled/reset at the start of the image and its value is polled at each clock cycle and transmitted into the datastream. The number of counts in each clock cycle has to be calculated by $\Delta C(t) = C(t+1)-C(t)$. The final column is an absolute timestamp, expressed as seconds since the device was powered on.

The data stream is processed by a supporting computer, which performs the following calculations: first, the position values are converted from interferometer resolution units to metric by multiplying them by the resolution of the interferometer ($p_{\mathrm{R},(x,y)}(t) = p_{\mathrm{M}(x,y)}(t) \times r_{(x,y)}$, being $r_{(x,y)} = 150$~pm in this case). As the interferometer position is recorded at the rising edge of the 4~kHz sampling clock, but the counts are recorded across the entire clock cycle, the position of the sample when these counts were acquired will be given by the average position between the current and the following rising edge of the sampling clock:

\begin{equation*}
p_{(x,y)}(t) = \frac{p_{\mathrm{R},(x,y)}(t) + p_{\mathrm{R},(x,y)}(t+1)}{2},
\end{equation*}

For each of these calculated positions, the difference in the detector counts between the current and the following rising edge of the sampling clock will be assigned. Using the same example data shown in Tab.~\ref{tab:datastream}, the processed data will be as shown in Tab.~\ref{tab:processed}. 

\begin{table}[!h]
    \centering
    \begin{tabular}{c | c | c}
        $p_x$ [\textmu m] & $p_y$ [\textmu m] & $\Delta$Counts [cts]\\
        \hline \hline
        ... & ... & ...\\
        \rowcolor{red}
        5228.7040 & 4854.1489 & 558\\
        \rowcolor{green}
        5228.7045 & 4854.1502 & 545\\
        \rowcolor{green}
        5228.7042 & 4854.1509 & 577\\
        \rowcolor{yellow}
        5228.7038 & 4854.1516 & 564\\
        \rowcolor{cyan}
        5228.7040 & 4854.1537 & 581\\
        \rowcolor{lightgray}
        5228.7045 & 4854.1552 & 575\\
        \rowcolor{lightgray}
        5228.7056 & 4854.1559 & 551\\
        ... & ... & ...\\
    \end{tabular}
    \caption{Example of a processed dataset. The rows are color-coded based on the pixel at which the data has been assigned (assuming a user-defined square binning size of 2~nm). Each row of the table corresponds to a sampling time of 250~\textmu s.}
    \label{tab:processed}
\end{table}

Now, the processed data is binned according to a user-defined spatial bin dimension. All the recorded photon counts that fall within a given bin are summed up, and the total time interval which the position remained in the corresponding bin is recorded as well, as shown in Tab.~\ref{tab:image} with the example data of Tab.~\ref{tab:processed}. This provides a collection of recorded counts in a user-defined grid, i.e. an image. However, due to stochastic variations in the sample position, there is no guarantee that the time in which the sample position is inside a given bin is uniform across the entire ROI (the PP and CV must also suffer similar issues and the most common strategy is to assume that the imperfections are not significant). Therefore, the recorded photon counts need to be normalized to the time interval of the defined bin, i.e. the per bin photon count rate needs to be displayed, as shown in Tab.~\ref{tab:image}. Note that the Poisson noise still needs to be calculated from the recorded photon counts.

\begin{table}[!h]
    \centering
    \begin{tabular}{c | c | c | c}
        Counts [cts] & Sample time [\textmu s] & Count rate [Mcts/s] & Poisson noise [\%]\\
        \hline \hline
        ... & ... & ... & ...\\
        \rowcolor{red}
        558 & 250 & 2.232 & 4.23\\
        \rowcolor{green}
        1122 & 500 & 2.244 & 2.98\\
        \rowcolor{yellow}
        564 & 250 & 2.26 & 4.21\\
        \rowcolor{cyan}
        581 & 250 & 2.324 & 4.15\\
        \rowcolor{lightgray}
        1126 & 500 & 2.252 & 2.98\\
        ... & ... & ... & ...\\
    \end{tabular}
    \caption{Example of a binned dataset using the same data as in Tab.~\ref{tab:datastream}. Each row of the table represents a pixel in the final reconstructed image, with the same color coding as in Tab.~\ref{tab:processed}. Note here that, due to the fact that the pixels of the reconstructed image do not have the same dwell time, the quantity displayed will be the count rate, and not the total recorded counts. However, the Poisson noise is still determined by the total recorded counts.}
    \label{tab:image}
\end{table}

The possibility to assign the recorded photon counts to the measured sample position allows for a correction of errors that arise from assuming a wrong sample position. This principle is shown in Fig.~\ref{fig:setup}(b), where data points are assigned to different pixels depending on whether the data was acquired in CV mode (i.e. all data acquired within a dwell time is assigned to one pixel), or using the supersampled method, where the data is re-assigned to the correct pixel after the measurement.

\begin{figure}
    \centering
    \includegraphics[width=0.4\textwidth]{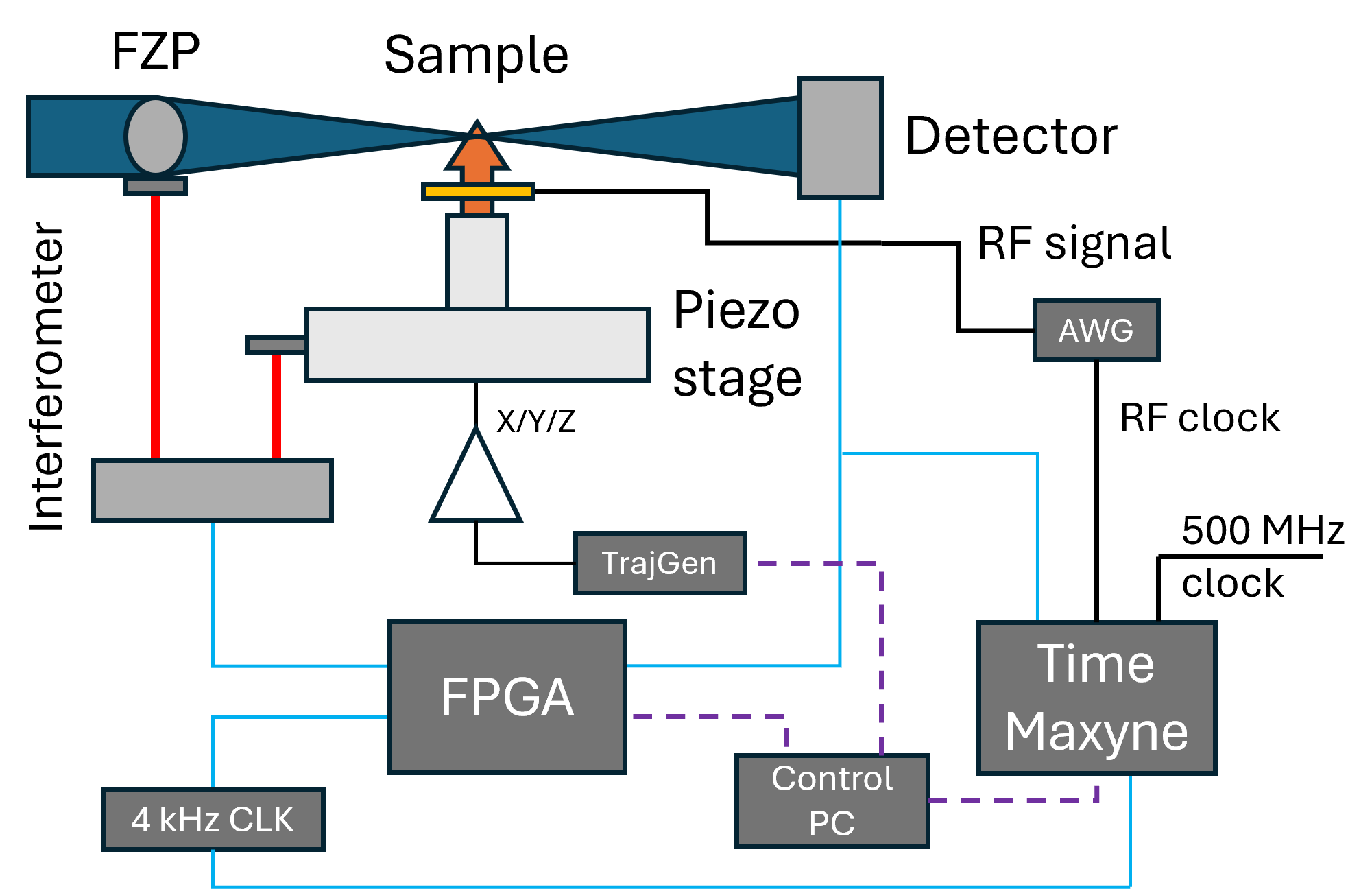}
    \caption{Sketch of the setup employed for the time-resolved measurements combined with supersampled microscopy. The \textit{TimeMaxyne} \cite{Weigand2022} setup is utilized for the acquisition of the time trace for each sampling point.}
    \label{fig:tr_setup}
\end{figure}

The supersample protocol described above can also be combined with the already-existing time-resolved (TR) STXM imaging setups \cite{Weigand2022, Puzic2010, Finizio2020}. This is of particular interest, as TR-STXM imaging is typically performed in PP mode, leading to significant image overheads (see Fig.~\ref{fig:overhead}). TR-STXM imaging relies on a generalization of the \textit{pump-probe} protocol and is described in detail elsewhere \cite{Weigand2022, Puzic2010, Finizio2020}. For the purpose of this work, the TR-STXM imaging protocol yields, for each pixel, a time series of the dynamical process under investigation. TR-STXM can be integrated with the supersampled imaging protocol by replacing the \textit{pixel clock} used in standard TR-STXM imaging to trigger the acquisition of the time series with the position sampling clock (in this case, the 4~kHz clock), therefore acquiring a time series for each cycle of the position sampling clock. The block diagram for this setup is shown in Fig. \ref{fig:tr_setup}. For the work described here, the TimeMaxyne setup of the Maxymus beamline \cite{Weigand2022} was utilized, which requires the 500~MHz master clock of the synchrotron light source to synchronize the excitation applied to the sample with the X-ray pulses used to probe the dynamical state of the sample. The resulting dataset is similar to the one shown in Tab. \ref{tab:datastream}, but with a time series replacing the total number of counts. The number of points in the time series will depend on the specific parameters of the TR scan. For the experiments shown here, we selected a time series with 7 points ($N = 7$ and $M = 1$ using the notation described in \cite{Weigand2022} - resulting in an excitation frequency of about 71.43~MHz).

\section{Discussion}

To demonstrate the functionality and performance of the concept described above, we imaged a 10~nm Siemens star. This measurement is suited for demonstration of reduction of the imaging overhead, compensation of vibration effects and positioning glitches, and high-resolution imaging ($< 10$~nm pixel size). The Siemens star was fabricated at PSI with the line-doubling technique \cite{Roesner2018}, and consists of SiO$_2$ spokes coated with a 10~nm Ir layer, whereas the capping layer was removed after fabrication. At its center, the distance between two adjacent Ir layers is of approximately 11~nm. The Siemens star was imaged with two different FZPs. The first FZP, which was used for the vibration compensation proof-of-concept measurement, is a 240~\textmu m diameter Au FZP with an outermost zone width of 35~nm. The second one, which was used for the proof-of-concept high-resolution imaging experiment, is a 240~\textmu m diameter Ir line-doubled FZP with an outermost zone width of 8.8~nm. Fabrication details of the high-resolution FZP can be found in \cite{Roesner2018, Roesner2020}. For all the measurements reported here, the beamline settings were selected to guarantee a coherent illumination of the FZP, allowing us to reach the theoretical resolution limit of the X-ray lens. All of the Siemens star images shown below were acquired at a photon energy of 1~keV.

The overhead reduction was verified by performing the same image sequence shown in Fig.~\ref{fig:overhead}. For the supersampled imaging method, the images were acquired in CV mode, reducing the nominal scan size along the \textit{x} axis to account for the stage acceleration and deceleration distances (i.e. $d = d_{\mathrm{ROI}}-d_\mathrm{acc} - d_\mathrm{decel}$, where $d_\mathrm{ROI}$ is the desired scan size and $d$ the scan size inserted in the Pixelator control software). As both the acceleration and deceleration regions contribute to the final image, a total overhead time of 10-15~\% was achieved with this method, independently of the desired pixel dwell time, and always lower than using either the PP or the CV modes (note here that the exact value of the overhead will still depend on parameters such as the number of lines in the image and the size of the ROI - but it will always be lowest with the supersampled imaging method). This is especially important at low dwell times, and will be of relevance for fast imaging experiments enabled by DLSRs, where the image overhead of the supersampled imaging method is significantly lower. Improvements on the positioning protocol may reduce this overhead further since the precision requirements are shifted from the relatively slow positioning feedback system to the metrology, allowing one to perform faster scanning. 

\begin{figure}[hbt]
    \includegraphics[width=0.4\textwidth]{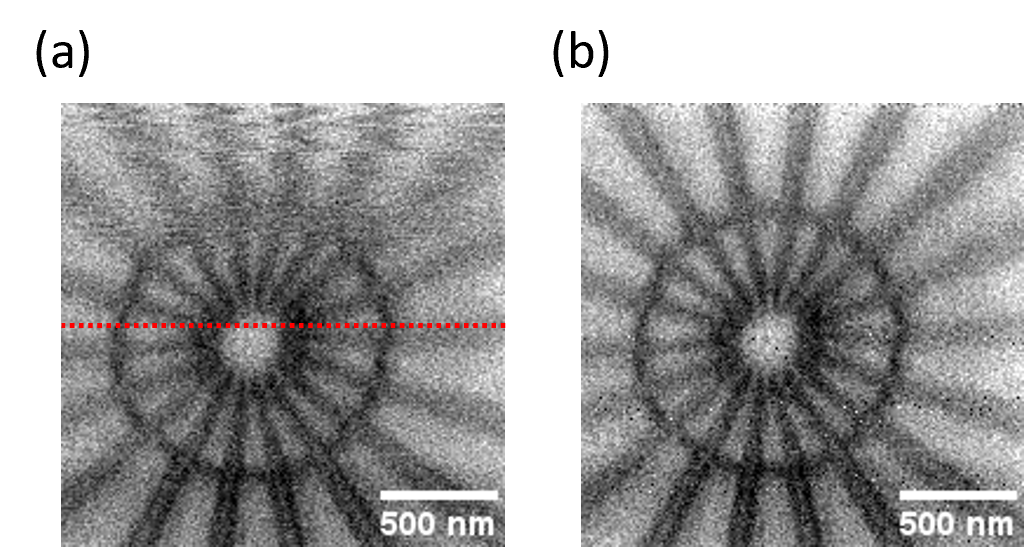}
    \caption{Proof-of-concept for vibration control. (a) STXM image acquired of a 10~nm Ir Siemens star acquired with the standard STXM setup. In the top half of the image, above the red line, a closed cycle cryostat cooler was turned on to demonstrate the effect of strong mechanical vibrations. (b) Same dataset as in (a), but reconstructed with the supersampling method. The reconstructed image shows no vibrations. Both images have a pixel size of 10~nm. Imaging performed with a 35~nm outermost zone FZP under diffraction-limited conditions. The effective dwell time is of 8~ms/px (Poisson noise of 1.3~\%, compared to a contrast of 2.6~\%).}
    \label{fig:vibrationControl}
\end{figure}

The proof-of-concept vibration compensation measurement is shown in Fig.~\ref{fig:vibrationControl}. In order to compare between the standard STXM imaging and the supersampled scanning microscopy method, an image was acquired in CV mode, while the same data was recorded in parallel using the PandABox. At half of the scan, a closed cycle cryostat cooler was switched on, generating significant mechanical vibrations that couple into the sample stage. Their effect are clearly visible in the standard CV STXM scan (Fig.~\ref{fig:vibrationControl}(a)). In stark contrast, the sample vibrations can be completely mitigated with the supersampled method (Fig.~\ref{fig:vibrationControl}(b)). Actually, strong sample vibrations have an advantageous side effect if scanning is performed in a line-at-once manner: high frequency vibrations enable a uniform sampling of the entire ROI, reducing variations in count rate errors (see the Supporting Information for additional details). Note, however, that only vibration modes where the distance between the sample and the interferometer mirror does not change can be compensated. High order vibration modes, and thermal drifts of the sample with respect to the interferometer mirror cannot be compensated in this way.

\begin{figure}[h]
    \includegraphics[width=0.3\textwidth]{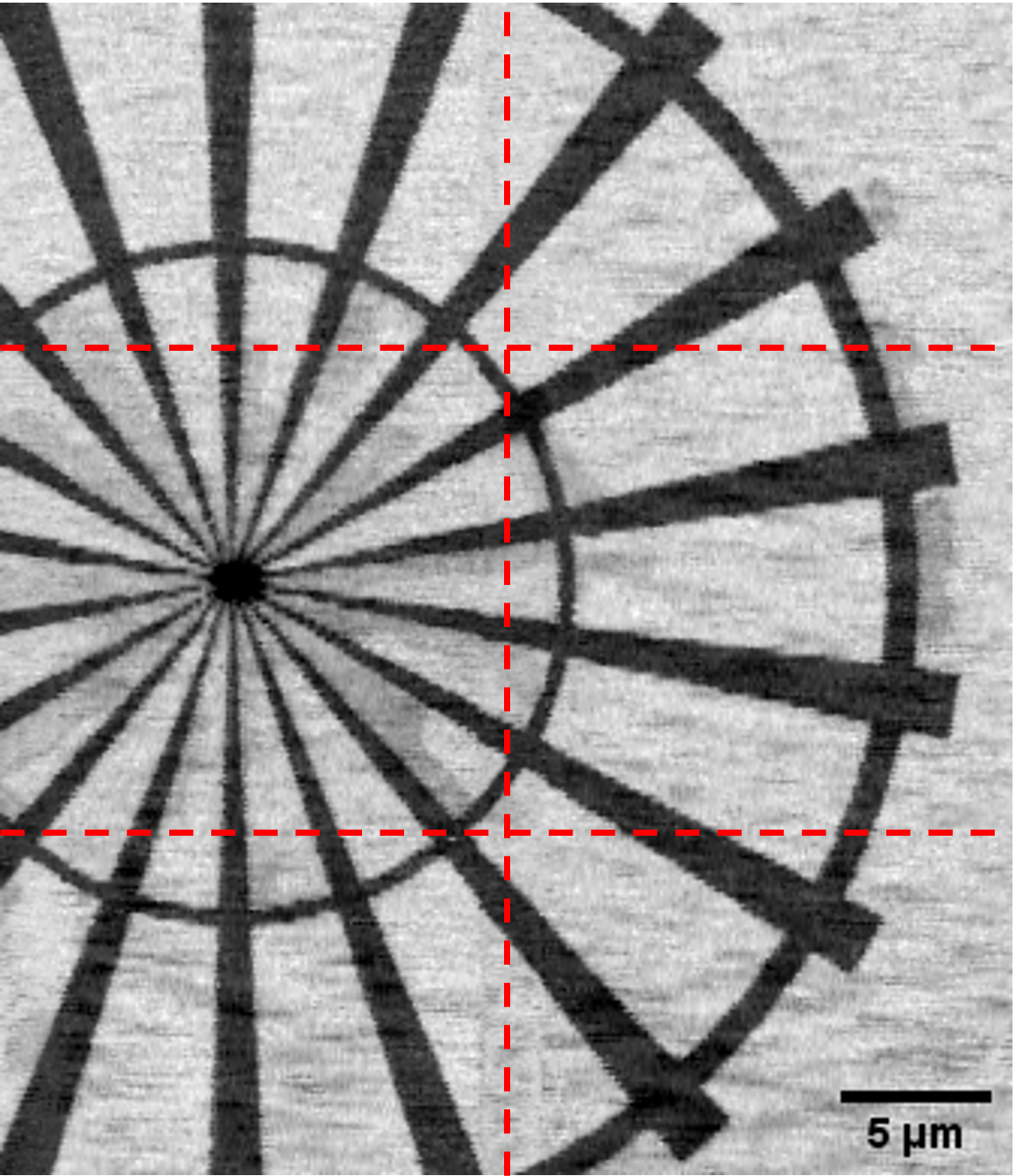}
    \caption{Large overview scan acquired with supersampled microscopy combined with tiling mode scanning. The borders of the single tiles are marked by the red dashed line. The image was reconstructed with a pixel size of 100 nm. Image acquired at the Maxymus endstation with an effective pixel dwell time of 0.5~ms/px (Poisson noise of 5.5~\%, compared to a contrast of 8.5~\%).}
    \label{fig:tilingMode}
\end{figure}

The typical range of the piezoelectric scanners used in STXM imaging is on the order of 10~\textmu m. Larger ROI scans are performed with the so-called \textit{tiling} mode, where several tiles smaller than the range of the piezoelectric scanner are individually scanned and stitched together. The coarse positioners of the endstation are used to move between the individual tiles. As the interferometric metrology is not limited by the range of the piezoelectric scanner, the supersampled microscopy method described here can be combined with the tiling mode, allowing one to acquire large overview scans as shown in Fig. \ref{fig:tilingMode}, where the data originating from different tiles could be successfully stitched to form a large overview image.

\begin{figure*}[hbt]
    \includegraphics[width=0.9\textwidth]{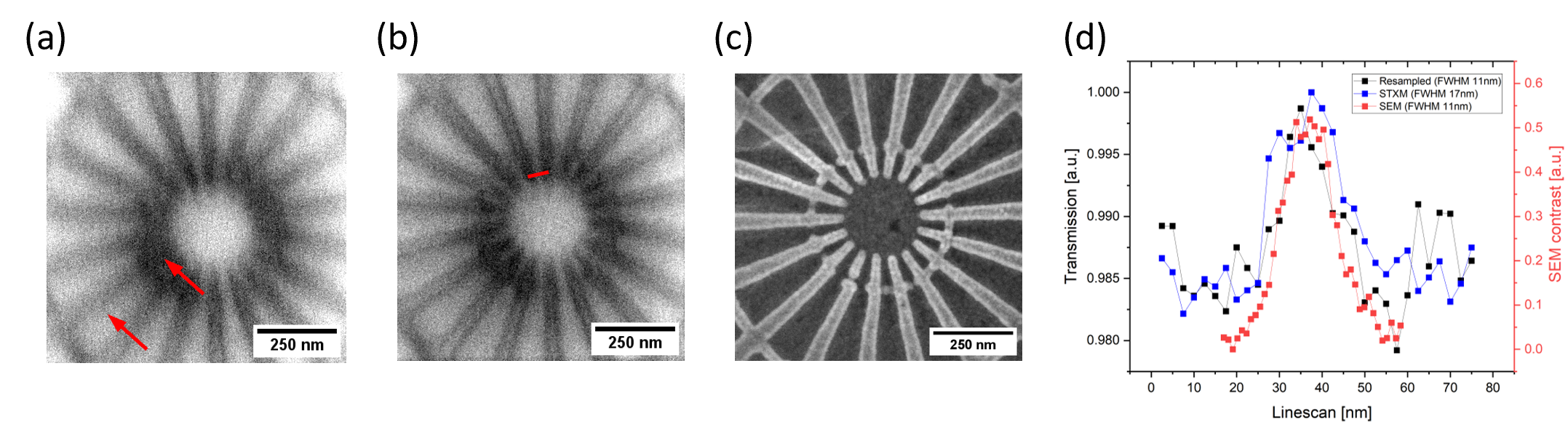}
    \caption{Proof-of-concept for high-resolution imaging. (a) High-resolution standard STXM image of the central area of a 10~nm Ir Siemens star. Several artifacts, marked by the red arrows, are visible in the image. A blur of the spokes in the vertical direction (caused by vibrations) is visible, and the center of the star cannot be fully resolved. (b) Same dataset as in (a), but evaluated from a supersampling stream. Using this method, the artifacts identified in (a) have been corrected, and the center of the star is well resolved. Both images have an effective pixel size of 2.5~nm. Imaging was performed with a 8.8~nm outermost zone FZP under diffraction-limited conditions. The effective dwell time is 38~ms/px (Poisson noise of 0.44~\%, compared to a contrast of 2.6~\%). (c) Scanning electron micrograph of the Siemens star imaged in (a-b). (d) Profile across the area marked in (b) derived from the standard STXM image, the supersampled reconstruction, and the SEM image. A significant reduction in the linewidth can be observed when using the supersampled reconstruction.}
    \label{fig:highRes}
\end{figure*}

To demonstrate the high-resolution imaging, we acquired an image with the aim to resolve the 10~nm linewidth at the center of the Siemens star. Here, the imaging was performed by repeatedly scanning a $900 \times 900$~nm$^2$ ROI around the center of the star following a meander pattern with a stage velocity of 5~\textmu m/s (CV scan with a nominal pixel size of 2.5~nm and a pixel dwell time of 0.5~ms). Due to the requirement of significantly narrowing down the secondary source size to guarantee coherent illumination conditions for the 8.8~nm outermost zone width FZP, a total pixel dwell time of about 38~ms was necessary to achieve the same ratio between contrast and Poisson noise (in practice, the ROI was scanned a total of 70 times, summing the recorded counts until the necessary signal-to-noise ratio was reached - this approach is used also for other techniques, such as ptychography imaging \cite{Deng2022}). To provide a fair comparison between the standard STXM imaging and the supersampled imaging method, particular attention was dedicated to the minimization of environmental vibration sources and drift sources: we performed the scans at night time when environmental influences are minimal and mounted the sample several hours before the measurement to provide enough settling time for thermal drifts and other related effects. Fig.~\ref{fig:highRes}(a) shows the resulting standard STXM image, where several imaging artifacts can be identified (marked by the red arrows in the figure). The marked artifacts (blur of the spokes and the insufficient spatial resolution in the center of the star along the \textit{x} axis) are attributed to the vibration of the sample along the \textit{y} axis (see Fig.~\ref{fig:vibrations}(a)). When the same data set is evaluated using the supersampled method, the spokes appear sharp and the center of the star can be resolved uniformly (Fig.~\ref{fig:highRes}(b)), yielding an image with a quality that almost reaches the superior resolution of scanning electron microscopy (Fig.~\ref{fig:highRes}(c)). This becomes particularly evident when deriving profiles at the center of the star, shown in Fig.~\ref{fig:highRes}(d), where a clear difference between the standard CV STXM and the supersampled imaging can be observed.

The supersampled microscopy protocol allows us to correct image distortions arising from scanning artifacts, mechanical vibrations (for common modes of the sample and interferometer mirrors), and sample drifts, as long as they are occurring on a longer timescale than the time required to scan the ROI at least once. However, image distortions arising from the optics (e.g., astigmatism of the FZP) cannot be corrected with supersampled imaging, as those distortions are inherently encoded in the probing beam.

This method is also applicable for spectromicroscopy imaging. As changing the photon energy is significantly slower than scanning the sample, a supersampled spectromicroscopy scan would be acquired as a series of individual supersampled images at different photon energies.

For the images reported above, the raw data rate transmitted from the PandABox was of about 228~kB/s. For the images displayed in Fig. \ref{fig:highRes}, an acquisition time of about 8-10 min was necessary, yielding raw data files on the order of 110-140 MB size. These data can, however, be efficiently compressed: for reference, the raw data for the images displayed in Fig. \ref{fig:highRes} can be compressed of a factor 10 by using the standard GZIP lossless deflate algorithm. However, even compressed, the raw data file size for this method will be significantly larger than for standard STXM images (using, again, the images shown in Fig. \ref{fig:highRes} as reference, the standard STXM raw image size is of about 1.5 MB), but these file sizes are easily manageable with modern computing infrastructure.

\begin{figure}[hbt]
    \includegraphics[width=0.4\textwidth]{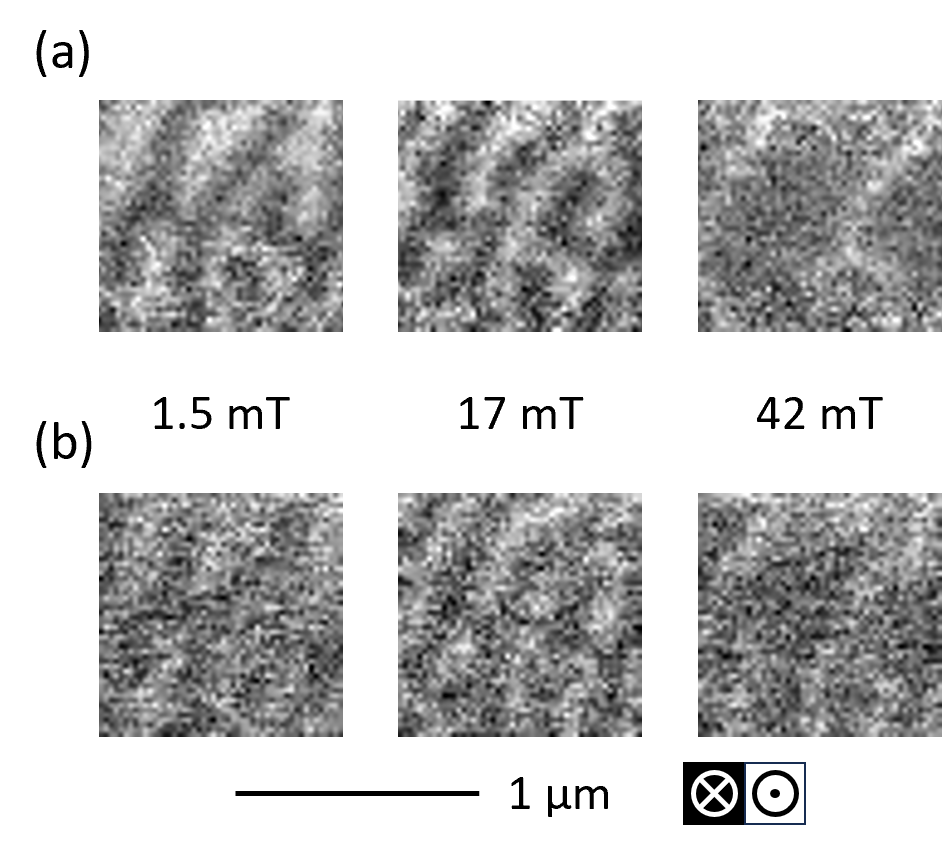}
    \caption{Single frames of a magnetic hysteresis loop in a SAF sample, cropped to a $1\times1$~\textmu m$^2$ area. (a) Acquisition performed using supersampled scanning microscopy. (b) Acquisition performed using standard CV STXM imaging. The direction of the magnetic contrast in the image (perpendicular magnetization) is indicated by the black/white bar. Each of the frames corresponds to a real time of 18~s. Effective pixel dwell time of 1.5 ms (Poisson noise of 1.5~\%, compared to a contrast of 2.5~\%). The full movie is shown in the Supporting Information.}
    \label{fig:movie}
\end{figure}

Up to now, we demonstrated the correction of vibrations and scanning glitches, as well as the reduction of overhead times. As all three issues are particularly problematic for imaging at low pixel dwell times, STXM has not been significantly utilized in the past for the imaging of mesoscopic dynamical processes, i.e. processes occurring on time scales of seconds to minutes: the relatively slow scanning velocity would give rise to rolling shutter artifacts. By enabling fast imaging at low pixel dwell times whilst maintaining image fidelity, our method significantly reduces the impact of the rolling shutter artifact, allowing for the imaging of mesoscopic dynamical processes at the second/minute timescales. As a proof-of-concept experiment, we recorded the magnetic hysteresis loop of a synthetic antiferromagnetic (SAF) multilayer superlattice, which is composed of 5 pairs of alternating layers of Co$_{68}$B$_{32}$ and Co$_{40}$Fe$_{40}$B$_{20}$ forming oppositely magnetized sublattices \cite{Barker2023}. Thanks to the different Co content of the two layers, the magnetization of the Co$_{68}$B$_{32}$ layer could be visualized by using circularly polarized X-rays tuned at the L$_3$ edge of Co (ca. 778~eV), allowing us to employ the X-ray magnetic circular dichroism (XMCD) effect as contrast mechanism \cite{Barker2023}. A 240~\textmu m diameter FZP with a 35~nm outermost zone width under coherent illumination conditions was utilized for this experiment. 

A ROI of 1 $\times 1$~\textmu m$^2$ (with an additional cumulative 0.3~\textmu m acceleration/deceleration distance) was repeatedly scanned in a meander pattern with 50 lines (20~nm/line) at a stage velocity of 10~\textmu m/s. A single ROI frame scan requires approximately 6~s to complete. The ROI frame scan was repeated 300 times, and the position-dependent photon count rate was recorded as described above. During the acquisition of this scan, the out-of-plane magnetic field applied to the SAF film, generated by means of a rotatable permanent magnet installed above the sample, was continuously modified from 0 to 50~mT at a velocity of about 1~mT every 6 scanned frames (i.e. 24~\textmu T/s). The recorded data was subdivided in 3 ROI-scans datasets. From each of them, an image of the magnetic configuration of the sample was reconstructed. This resulted in a 100 frame movie of the evolution of the magnetic state of the SAF film under an applied out-of-plane magnetic field from 0 to 50~mT with a 0.5~mT step/frame. A choice of recorded frames (the full movie is shown in the Supporting Information - note that, as the acceleration and deceleration regions are also recorded in the supersampled image, the ROI will be of $1.3 \times 1$~\textmu m$^2$, while the regular STXM image will have a $1 \times 1$~\textmu m$^2$ ROI) is shown in Fig.~\ref{fig:movie}(a), where they can be compared with the same frames acquired with the standard CV STXM setup (Fig.~\ref{fig:movie}(b)). In the latter, several positioning artifacts can be observed. Each frame corresponds to a real time step of 18~s, giving a real time of 30 minutes for the movie. This timestep allows us to directly image (i.e. not stroboscopic imaging) dynamic processes occurring at the minute timescale, whilst maintaining a high image quality and fidelity. It has to be noted that the limiting factor for the timestep used in the recorded movie was not the velocity of the piezoelectric stage (a maximum stage velocity of 300~\textmu m/s can be achieved) but rather the acquired image statistics (i.e. Poisson noise-limited imaging). As PolLux is a bending magnet beamline, the photon flux at the detector is limited to about 1~MPh/s \cite{Raabe2008}, which results in a Poisson noise of about 1.5~\% for each of the frames shown in Fig.~\ref{fig:movie}, compared to an XMCD contrast of about 2.5~\% for this sample. Thanks to the significantly higher coherent photon flux delivered to the endstation \cite{Streun2018}, an undulator-based beamline at a DLSR source will easily overcome this limitation, allowing for the acquisition of (non-reproducible) dynamics in the second and sub-second timescales. This is of interest for many applications, such as, e.g., the investigation of nucleation processes and phase transitions \cite{Alpert2022}.

\begin{figure}
    \centering
    \includegraphics[width=0.25\textwidth]{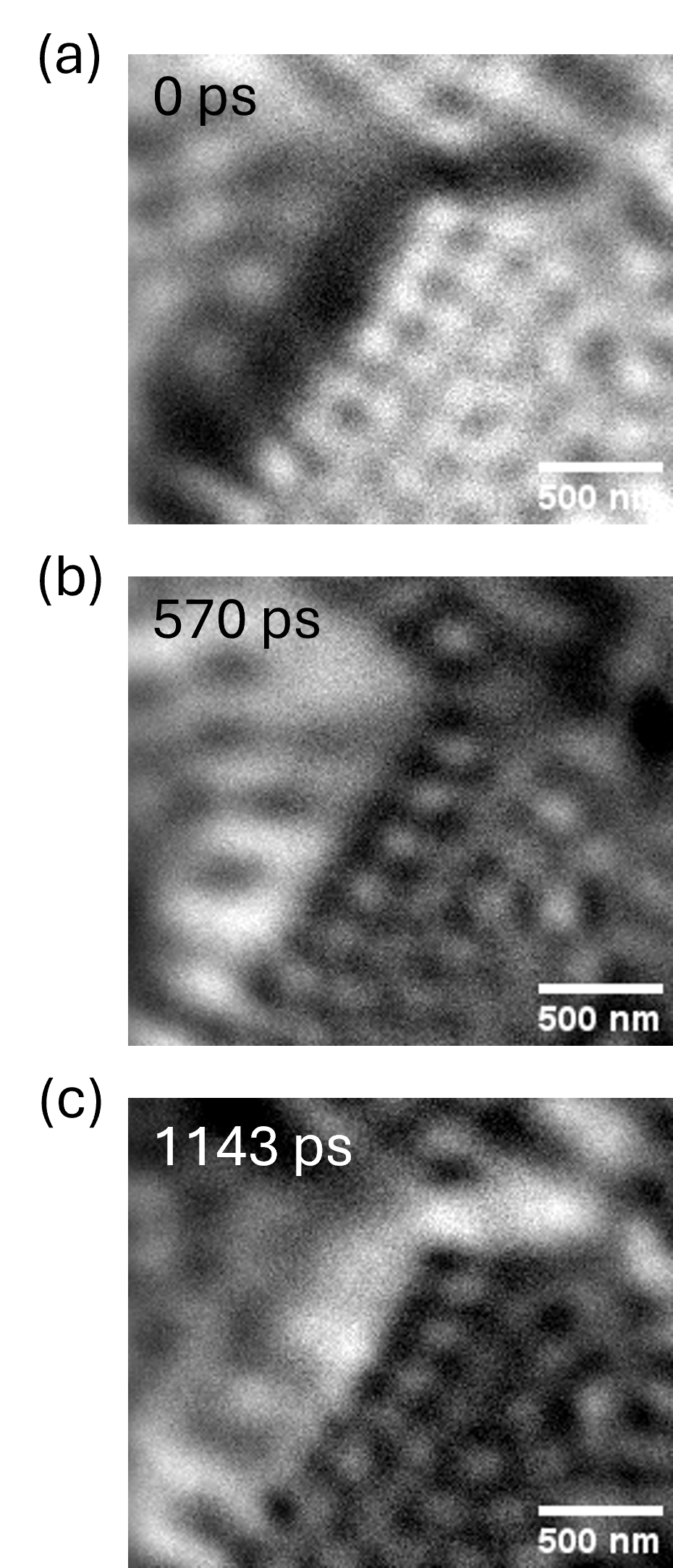}
    \caption{Single frames of a TR image acquired combining the pump-probe and supersampled imaging protocols. The frames have a pixel size of 10~nm. Image acquired at the Maxymus endstation with an effective pixel dwell time of 4~ms/px for each frame (Poisson noise of 0.1~\% per frame, compared to a contrast of 5~\%).}
    \label{fig:tr_scan}
\end{figure}

Finally, we demonstrated the combination of the supersampled imaging method with pump-probe TR-STXM imaging. For this proof-of-concept, we imaged the spin-wave dynamics of a 112~nm thick perpendicularly magnetized polycrystalline Yttrium Iron Garnet (YIG) sample grown by sputter deposition directly onto a Si$_3$N$_4$ membrane. The spin-wave dynamics were excited by injecting a 71.43~MHz ($N = 7$ and $M = 1$ according to the notation in \cite{Weigand2022}) RF signal across a 1~\textmu m wide, 200~nm thick, patterned Cu stripline, generating an oscillating magnetic field through the Oersted effect. The magnetization dynamics were visualized by using circularly-polarized X-rays tuned to the L$_{3\mathrm{b}}$ edge of Fe (ca. 709.3~eV). The reconstruction of the final images was performed as described above for each point in the time series (i.e., in this case, yielding 7 images). As usual for TR-STXM imaging, each of the 7 frames of the TR image was divided by the average of all 7 frames, to produce a normalized image where changes to the magnetic configuration are easier to visualize \cite{Weigand2022}. A selection of frames from the resulting scan, reconstructed with a 10~nm pixel size, are shown in Fig. \ref{fig:tr_scan}. The full time-resolved movie is shown in the Supporting Information.

\section{Conclusions}

In conclusion, we describe a new method for the acquisition of STXM images, called supersampled scanning microscopy. This concept is based on the measurement of the sample position and the detected photon counts at a high rate (faster than the mechanical vibrations), and subsequently binning of the data according to the recorded positions. An X-ray microscopy image can be obtained in this way, where positioning artifacts and mechanical vibrations are fully compensated. As additional advantages, the technique allows for a relaxation of the positioning precision in the scanning, significantly reducing imaging overheads if compared to the standard STXM PP and CV imaging protocols, whilst maintaining the possibility to acquire high-resolution images. Furthermore, the technique allows one to reduce the timescale of the rolling shutter artifact that is typical of scanning microscopy techniques, permitting the acquisition of mesoscopic dynamic movies at timescales limited by the available photon fluxes and scanning velocity of the piezoelectric stage. The performance of the technique was demonstrated by proof-of-concept experiments on the reduction of the imaging overhead, compensation of vibrations and positioning glitches, high-resolution imaging, and the acquisition of mesoscopic dynamic movies. Finally, the supersampled scanning microscopy technique can also be integrated with TR imaging, allowing one to exploit the improvement in imaging overhead and in the mitigation of the influence of mechanical vibrations also for TR imaging. A proof-of-concept experiment where the spin-wave dynamics in a perpendicularly magnetized YIG sample were imaged, was shown. This imaging protocol is currently being integrated into the Pixelator control system of the PolLux and Maxymus endstations \cite{Watts2018}, with the aim to include live visualization and automatic processing of the image, including e.g., the possibility to display the live signal-to-noise ratio (expressed as contrast vs. Poisson noise). It will be made available for regular user operation. The position sampling frequency will also be increased to 10~kHz once live processing of the image is available.

\begin{acknowledgement}

This work was performed at the PolLux (X07DA) beamline of the Swiss Light Source, Paul Scherrer Institut, Villigen, Switzerland and at the Maxymus beamline of the Bessy II synchrotron light source, Helmholtz Zentrum Berlin, Germany. The PolLux endstation was financed by the German Bundesministerium f\"{u}r Bildung und Forschung through ErUM-Pro contracts 05K16WED, 05K19WE2, and 05K22WE2. This project was financed by the Swiss National Science Foundation under grant agreement No. CRSK-2\_227418. T.A.B. acknowledges funding from the European Regional Development Fund. The authors thank Christopher Barker, Eloi Haltz, Christopher Marrows (University of Leeds) for providing the SAF sample, and Karthink Srinivasan, Bethanie J H Stadler (University of Minnesota), and Kai Litzius (University of Augsburg) for providing the YIG sample.

\end{acknowledgement}

\begin{suppinfo}

\begin{itemize}
    \item Details of the setup of the PandABox FPGA, of the distribution of pixel dwell times resulting from the rebinning process, together with additional high-resolution images and with a display of the effect of mechanical vibrations on the uniformity of the sampling. (PDF)
    \item Video displaying the response of the SAF sample to a varying magnetic field, from which the frames shown in Fig. \ref{fig:movie}(a) are shown. This video displays the data processed with the supersampled scanning microscopy method. (SAF\_Supersample.avi)
    \item Video displaying the response of the SAF sample to a varying magnetic field, from which the frames shown in Fig. \ref{fig:movie}(b) are shown. This video displays the data acquired with standard CV STXM imaging. (SAF\_STXM.avi)
    \item Video displaying the spin-wave dynamics in a YIG sample excited by a 71.43~MHz RF signal acquired with the supersampled scanning microscopy method, from which the frames shown in Fig. \ref{fig:tr_scan} are shown. (TR\_STXM.avi)
\end{itemize}

\end{suppinfo}

\bibliography{bibliography}

\end{document}